\documentclass{elsart}
\usepackage{graphicx}
\usepackage{amsmath}
\usepackage{amssymb}

\begin{document}

\begin{frontmatter}

\title{
Quasi two-dimensional carriers
in dilute-magnetic-semiconductor quantum wells
under in-plane magnetic field}

\author[address]{Constantinos~Simserides\thanksref{thank-s}}, and
\author[address]{Iosif~Galanakis}

\address[address]{University of Patras, Materials Science Department,
Panepistimiopolis, Rio, GR-26504, Patras, Greece}

\thanks[thank-s]{Corresponding author. E-mail: simserides@upatras.gr}

\begin{abstract}
Due to the competition between spatial and magnetic confinement,
the density of states of a quasi two-dimensional system
deviates from the ideal step-like form
both quantitatively and qualitatively.
We study how this affects
the spin-subband populations and the spin-polarization
as functions of the temperature, $T$,
and the in-plane magnetic field, $B$,
for narrow to wide
dilute-magnetic-semiconductor quantum wells.
We focus on the quantum well width,
the magnitude of the spin-spin exchange interaction,
and the sheet carrier concentration dependence.
We look for ranges where the system
is completely spin-polarized.
Increasing $T$, the carrier spin-splitting, $U_{o\sigma}$, decreases,
while
increasing $B$, $U_{o\sigma}$ increases.
Moreover, due to the density of states modification,
all energetically higher subbands become gradually depopulated.
\end{abstract}

\end{frontmatter}

\section{Introduction}
The progress in understanding of transition-metal-doped semiconductors
has been impressive
\cite{jungwirth_et_al_review-2006,dietl-ohno-review-2006,macdonald-et-al-review-2005}.
New phenomena and applications have been discovered,
in transition metal doped III-V or II-VI compounds
including quasi two-dimensional systems \cite{simserides-2007,kechrakos-et-al-2005}
where wave function engineering may play a substantial role e.g.
increase the ferromagnetic transition temperature.

An in-plane magnetic field
distorts the density of states (DOS) \cite{lyo-1994,simserides-1999}
of a quasi two-dimensional system because the spatial and the magnetic confinement compete.
The energy dispersion in the $xz$-plane has the form
$E_{i,\sigma}(k_x)$, where $i$ is the subband index, $\sigma$ is the spin,
$k_x$ the in-plane wave vector perpendicular to the in-plane magnetic field,
$B$ (applied along $y$), and $z$ is the growth axis.
The envelope functions along $z$ depend on $k_x$ i.e.,
$\psi_{i,\sigma,k_x,k_y}({\bf r})
\propto \phi_{i,\sigma,k_x}(z) e^{i k_x x} e^{i k_y y}$.
This modification has been realized in
transport \cite{transport} and
photoluminescence \cite{PL} studies, as well as
in the detection of plasmons in QWs \cite{plasmons}.
A fluctuation of the magnetization
in dilute-magnetic-semiconductor (DMS) structures
in cases of strong competition between spatial and magnetic confinement
has been predicted at low enough temperatures \cite{simserides-prb-2004}
and a compact DOS formula holding for any type of interplay
between spatial and magnetic confinement exists \cite{simserides-prb-2004}:

\begin{equation}\label{dos}
\rho({\mathcal E}) = \frac {A \sqrt{2m^*}}{4 \pi^2 \hbar}
\sum_{i,\sigma} \int_{-\infty}^{+\infty} \! dk_x
\frac{\Theta({\mathcal E}-E_{i,\sigma}(k_x))}
{ \sqrt {{\mathcal E}-E_{i,\sigma}(k_x)} }.
\end{equation}

\noindent $\Theta$ is the step function, $A$ is the $xy$ area
of the structure, $m^*$ is the effective mass.
Generally, $E_{i,\sigma}(k_x)$ must be self-consistently calculated
\cite{simserides-1999,transport,PL,simserides-prb-2004}.
The $k_x$-dependence in Eq.~(\ref{dos}) increases
the numerical cost by a factor of $10^2-10^3$ in many cases.
For this reason, in the past, the $k_x$-dependence has been quite often ignored,
although this is only justified for narrow QWs.
With the existing computational power,
such a compromise is not any more necessary.
In the limit $B \to 0$, the DOS retains
the occasionally stereotypic staircase shape
with the {\it ideal} step
$\frac {1}{2} \frac {m^* A}{\pi\hbar^2}$ for each spin.
The DOS modification significantly affects the physical properties
and specifically the spin-subband populations and spin polarization
in DMS quantum wells (QWs) \cite{simserides-2007}.
For completeness, we notice that Eq.~(\ref{dos}) ignores disorder
which -with the current epitaxial techniques-
is important when the concentration of paramagnetic ions (e.g. Mn$^{+2}$ ) is high.

Here we briefly describe
how the above mentioned DOS determines the spin-subband populations and
the spin-polarization as functions of $B$ and the temperature, $T$,
for DMS single QWs giving a few examples.
Calculations for double QWs
will hopefully be published  in the future.
For narrow QWs, it has been shown \cite{simserides-prb-2004} that
the DOS is an almost ``perfect staircase''
with steps increasing only a few percent
relatively to the ideal 2DEG step.
In such a case, at very low $T$,
a completely spin-polarized system can also be achieved \cite{simserides-prb-2004}.

\section{A few equations}
\begin{equation}\label{spin-splitting}
U_{o\sigma} =
\frac {g^*m^*}{2m_e} \hbar \omega_c -
y N_0 J_{sp-d} S B_{S}(\xi) = \alpha + \beta .
\end{equation}

\noindent is the electron spin-splitting.
$\hbar \omega_c$ is the cyclotron gap,
$\alpha = \alpha (B)$
describes the Zeeman coupling between
the spin of the itinerant carrier and
the magnetic field,
while $\beta = \beta (B,T)$ expresses
the exchange interaction between
the spins of the Mn$^{+2}$ cations and
the spin of the itinerant carrier.
$g^*$ is the g-factor of the itinerant carrier.
$y$ is the molecular fraction of Mn.
$N_0$ is the concentration of cations.
$J_{sp-d}$ is the coupling strength due to the spin-spin exchange interaction
between the d electrons of the Mn$^{+2}$ cations
and the s- or p-band electrons,
and it is negative for conduction band electrons.
The factor $S B_{S}(\xi)$ represents the spin polarization of the Mn$^{+2}$ cations.
The spin of the Mn$^{+2}$ cation is $S =$ 5/2.
$B_{S}(\xi)$ is the standard Brillouin function.
Such a simplified Brillouin-function approach
is quite common when dealing with
quasi two-dimensional systems.
This way, the spin-orbit coupling is not taken into account.
This is certainly a simplification, since increasing $T$,
the magnetization of the magnetic ions competes with spin-orbit coupling.

\begin{equation}\label{xi}
\xi=\frac{g_{Mn}\mu_BSB
-J_{sp-d}S \frac{n_{down}-n_{up}}{2}}{k_BT}.
\end{equation}

\noindent $k_B$ is the Boltzmann constant. $\mu_B$ is the Bohr magneton.
$g_{Mn} $ is the $g$ factor of Mn.
$n_{down}$ and $n_{up}$ are
the spin-down and spin-up concentrations measured e.g. in cm$^{-3}$,
while $N_{s,down}$ and $N_{s,up}$ used below are
the spin-down and spin-up two-dimensional (sheet) concentrations
measured e.g. in cm$^{-2}$.
In Eq.~\ref{xi} (and only there) we approximate
$n_{down}-n_{up} \approx (N_{s,down} - N_{s,up}) / L$,
where $L$ is the QW width.
The first term in the numerator of Eq.~\ref{xi} represents
the contribution of the Zeeman coupling between the localized spin
and the magnetic field.
The second term in the numerator of Eq.~\ref{xi}
(sometimes called ``feedback mechanism'') represents
the kinetic exchange contribution
which -in principle- can induce spontaneous spin-polarization
i.e. in the absence of an external magnetic field.
Notice that $n_{down} - n_{up}$
is positive for conduction band electrons.
Finally, for conduction band electrons,
the spin polarization is

\begin{equation}\label{zeta}
\zeta = \frac {N_{s,down}-N_{s,up}}{N_{s}}.
\end{equation}

\noindent $N_{s} = N_{s,down} + N_{s,up}$ is the free carrier
two-dimensional (sheet) concentration.

$B$ and $T$ influence
the spin polarization in an opposite manner.
Moreover, for each type of spin population,
the in-plane magnetic field
-via the distortion of the DOS-
redistributes the electrons between the subbands i.e.,
all excited states become
gradually depopulated \cite{simserides-2007}.
Thus, the spin polarization can be tuned
by varying the temperature and the magnetic field.

\section{A few results and some discussion}
\begin{figure}[h] 
\begin{center}\leavevmode
\includegraphics[width=0.45\linewidth]{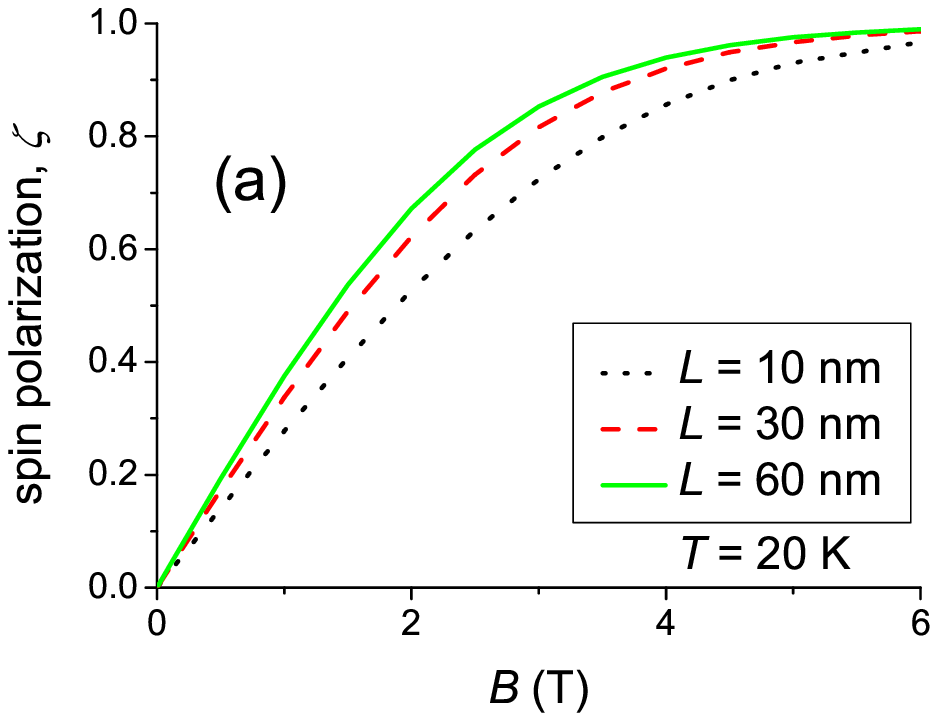}
\includegraphics[width=0.45\linewidth]{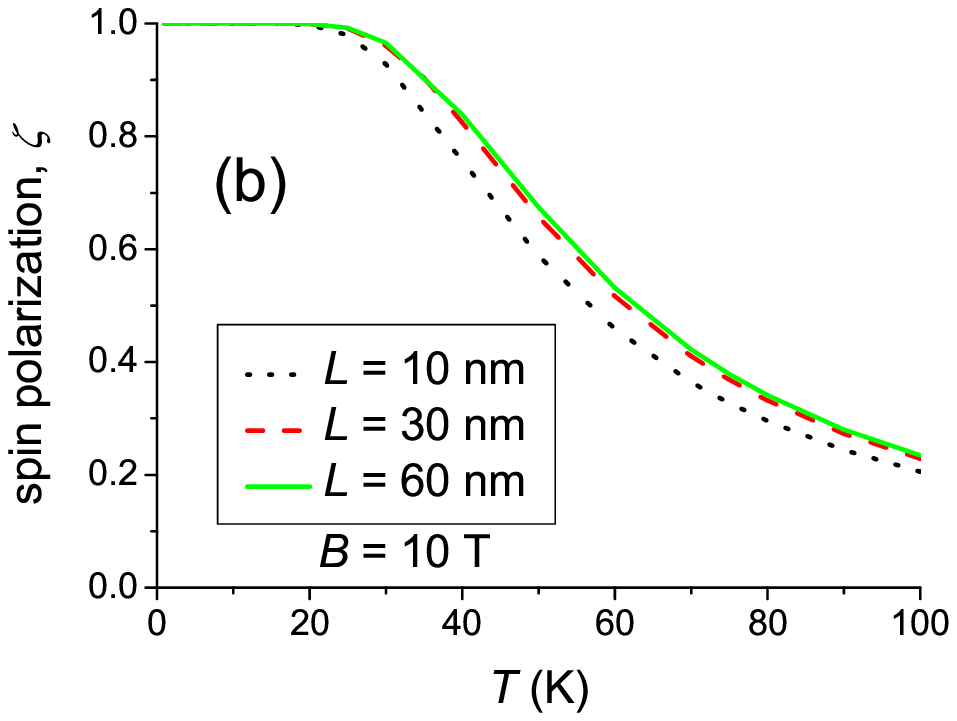}
\caption{The spin polarization, $\zeta$, tuned by varying:
(a) the in-plane magnetic field, $B$ ($T = $ 20 K), and
(b) the temperature, $T$ ($B =$ 10 T),
for different well widths, $L = $ 10 nm, 30 nm, and 60 nm.
$- J_{sp-d} = 12 \times 10^{-3}$ eV nm$^3$.}
\label{zeta_of_B_and_T_10nm_30nm_60nm}
\end{center}
\end{figure}

Details on the material parameters used here
can be found elsewhere \cite{simserides-2007}.
Figure \ref{zeta_of_B_and_T_10nm_30nm_60nm} depicts
the spin polarization tuned by varying
the parallel magnetic field and the temperature,
for different choices of the well width.
$- J_{sp-d} = 12 \times 10^{-3}$ eV nm$^3$,
while $N_s =$ 1.566 $\times$ 10$^{11}$ cm$^{-2}$.
Because of the DOS modification \cite{simserides-prb-2004},
resulting in different distribution of electrons
among the spin-subbands \cite{simserides-2007},
we witness a clear increase of $\zeta = \zeta (L)$, i.e.
$\zeta (L = $ 60 nm) $ >  \zeta ( L = $ 30 nm) $ > \zeta ( L = $ 10 nm).
For $B =$ 0, $\zeta$ vanishes,
i.e. there is no spontaneous spin polarization phase
due to the tiny ``feedback mechanism''
for this choice of material parameters.
Detailed illustrations of the effect of an in-plane magnetic field
on the energy dispersion as well as on the density of states,
for different well widths,
can be found elsewhere \cite{simserides-prb-2004,simserides-physicae-2004}.

\begin{figure}
\begin{center}\leavevmode
\includegraphics[width=0.75\linewidth]{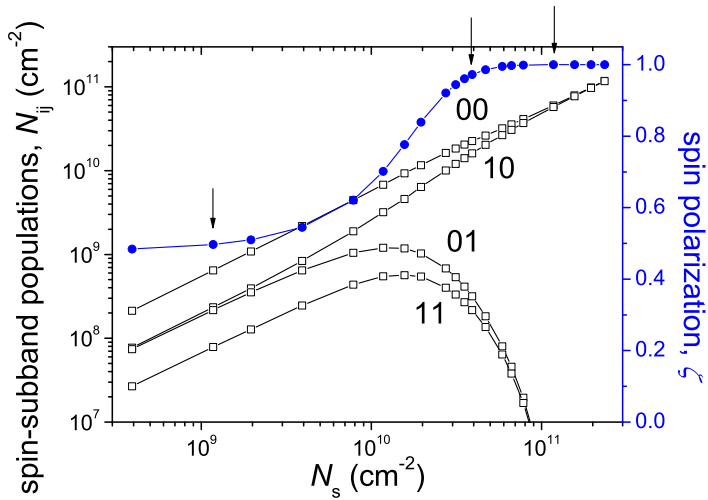}
\caption{The spin-subband populations, $N_{ij}$ and
the spin polarization, $\zeta$ (full symbols),
tuned by varying the sheet carrier concentration, $N_s$,
for $L = $ 60 nm, $T = $ 20 K and $B =$ 0.01 T,
using  $J = 12 \times 10^{-1}$ eV nm$^3$.
The arrows indicate $N_s$ values
where we compare with $B =$ 0.0001 T in the text.}
\label{ssp_and_zeta_of_Ns_FF}
\end{center}
\end{figure}

\begin{figure}[h] 
\begin{center}\leavevmode
\includegraphics[width=0.75\linewidth]{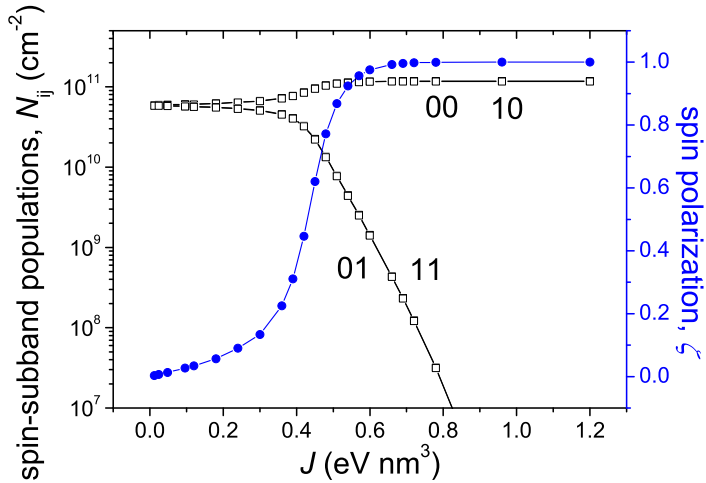}
\caption{The spin-subband populations, $N_{ij}$ and
the spin polarization, $\zeta$ (full symbols),
as a function of $J$.
$T =$ 20 K, $B =$ 0.01 T, $L =$ 60 nm, $N_s =$ 2.349 $\times$ 10$^{11}$ cm$^{-2}$.
}
\label{J-dependence}
\end{center}
\end{figure}

The influence of $N_s$ on
the spin-subband populations and the spin polarization
for different values of
the magnitude of the spin-spin exchange interaction,
$J$ is examined below.
In a heterostructure with higher $N_s$
we may require smaller values of $J$ in order to completely spin-polarize carriers.
Modifying $J$, we have explored the $N_s$ influence.
For $J = 12 \times 10^{-2}$ eV nm$^3$
there is a very small influence of $N_s$ on $\zeta$.
The situation changes using $J = 12 \times 10^{-1}$ eV nm$^3$.
Figure \ref{ssp_and_zeta_of_Ns_FF} shows
$N_{ij}$ and $\zeta$
tuned by varying $N_s$
for $L = $ 60 nm, $T = $ 20 K and $B =$ 0.01 T,
using  $J = 12 \times 10^{-1}$ eV nm$^3$.
The pair $ij$ is defined in the following manner:
00 symbolizes the ground-state spin-down-subband,
10            the 1st excited spin-down-subband,
01            the ground-state spin-up-subband, and finally
11 symbolizes the 1st excited spin-up-subband.
Increase of $N_s$
from $\approx$ 1.0 $\times$ 10$^{9}$ cm$^{-2}$
  to $\approx$ 1.0 $\times$ 10$^{11}$ cm$^{-2}$
is sufficient to completely spin-polarize carriers.
This is purely due to the ``feedback mechanism''
stemming from the difference between the populations of
spin-down and spin-up carriers.
If we decrease $B$ from 0.01 T to 0.0001 T, then e.g.
(a) for $N_s =$ 1.175 $\times$ 10$^{9}$ cm$^{-2}$,
$\zeta$ changes from 0.497 to 0.005,
(b) for $N_s =$ 3.917 $\times$ 10$^{10}$ cm$^{-2}$,
$\zeta$ changes from 0.973 to 0.909,
however,
(c) for $N_s =$ 1.175 $\times$ 10$^{11}$ cm$^{-2}$,
$\zeta$ remains 1.
$N_{ij}$ and $\zeta$,
as a function of $J$ are depicted in Fig.~\ref{J-dependence}.
$T =$ 20 K, $B =$ 0.01 T, $L =$ 60 nm, $N_s =$ 2.349 $\times$ 10$^{11}$ cm$^{-2}$.
Due to the small values of $B$, $N_{10} \approx N_{00}$ and $N_{11} \approx N_{01}$.

\section{Conclusion}
We have studied the spin-subband-populations and the spin-polarization
of quasi two-dimensional carriers
in dilute-magnetic-semiconductor single quantum wells
under the influence of
an in-plane magnetic field.
The proper density of states was used for the first time,
incorporating the dependence on the in-plane wave vector
perpendicular to the in-plane magnetic field.
We have examined a range of parameters,
focusing on the quantum well width,
the magnitude of the spin-spin exchange interaction,
and the sheet carrier concentration.
We have shown that varying these parameters
we can manipulate the spin-polarization,
inducing spontaneous (i.e. for $B \to $ 0) spin-polarization.

\end{document}